\documentclass[aps,prc,preprint,floatfix,superscriptaddress,showpacs,amsfonts,amssymb,amsmath]{revtex4}
\usepackage{graphicx}
\usepackage{rotating}

\begin{document}
\title{Comment on "$^{138}$La-$^{138}$Ce-$^{136}$Ce nuclear cosmochronometer of the supernova neutrino process"}

\author{P.~von Neumann-Cosel}\email{vnc@ikp.tu-darmstadt.de}
\affiliation{Institut f\"ur Kernphysik, Technische Universit\"at
Darmstadt, 64289 Darmstadt, Germany}
\author{A.~Richter}
\affiliation{Institut f\"ur Kernphysik, Technische Universit\"at
Darmstadt, 64289 Darmstadt, Germany}\affiliation{ECT*, Villa
Tambosi, I-38050 Villazano (Trento), Italy}
\author{A.~Byelikov}\altaffiliation{Present address: AREVA NP GmbH,
63067 Offenbach, Germany} \affiliation{Institut f\"ur Kernphysik,
Technische Universit\"at Darmstadt, 64289 Darmstadt, Germany}

\date{\today}

\begin{abstract}
The nuclear chosmochronometer suggested by Hayakawa et al. [Phys.~
Rev.~C~{\bf 77},~065802~(2008)] based on the
$^{138}$La-$^{138}$Ce-$^{136}$Ce abundance ratio in presolar grains
would be affected by the existence of a hitherto unknown low-energy
$1^+$ state in $^{138}$La. Results of a recent high-resolution study
of the $^{138}$Ba($^3$He,$t$) reaction under kinematics selectively
populating $1^+$ states in $^{138}$La through Gamow-Teller
transitions provides strong evidence against the existence of such a
hypothetical state.
\end{abstract}


\maketitle


Hayakawa {\it et al.}~\cite{hay08} proposed a new cosmochronometer
based on the $^{138}$La-$^{138}$Ce-$^{136}$Ce abundance ratios in
presolar grains. It utilizes the different dominant production
mechanisms of $^{138}$La and $^{136,138}$Ce: while all result from
explosive nucleosynthesis in type II supernovae, the former nuclide
is a product of the $\nu$ process through charged-current reactions
\cite{gor01,heg05,bye07a} while the latter nuclides are synthesized
in the $p$ process \cite{arn03}. One possible pitfall for the scheme
(and also for the $\nu$ process origin of $^{138}$La) would be the
existence of a low-lying $1^+$ state in $^{138}$La allowing electron
capture (EC) decay to the stable $^{138}$Ba competing with the
highly hindered $\gamma$ decay to the $5^+$ ground state. In fact,
the final sentence of the paper reads {\it "We present that the
energy of the lowest $1^+$ state may affect the chronometer
performance and the $\nu$ process origin of $^{138}$La, and its
measurement is desired."} It is the purpose of this comment to
report on recent experiments which allow to constrain the existence
of such an EC branch.

The problem discussed in \cite{hay08} would be a $1^+$ state in
$^{138}$La with an excitation energy below the presently known
\cite{son03} lowest excited state ($E_x = 72$ keV, $J^\pi = 2^+$).
Such a state could be populated in the ($\nu_e,e)$ reaction either
directly or by population of higher-lying $1^+$ states and
subsequent $\gamma$ decay, since Gamow-Teller (GT) transitions are
expected to dominate the reaction cross section \cite{heg05}. For
such a level EC decay to the $^{138}$Ba ground state (g.s.) could be
competitive with $\gamma$ decay because of the large transition
multipolarity ($E4$) of the latter. In principle, also $\beta-$
decay to the g.s.\ of $^{138}$Ce is possible. However, the smaller
$Q$-value limits its branching ratio to a few percent of the EC
decay.

Recently, we have performed a study of the
$^{138}$Ba($^3$He,$t$)$^{138}$La reaction at the Research Center for
Nuclear Physics, Osaka, Japan, at an incident energy of 140
MeV/nucleon and under a scattering angle of $0^\circ$ \cite{bye07b}.
A brief account of the work was given in \cite{bye07a}. Such
experiments can be performed with excellent energy resolution
reaching values $\Delta E / E \simeq 10^{-5}$ in heavy nuclei
\cite{kal06}. At angles close to $0^\circ$ this reaction selectively
excites GT transitions. It is thus a spectroscopic tool to
investigate $1^+$ states in $^{138}$La. Furthermore, GT matrix
elements can be extracted from the data utilizing the procedures
discussed in \cite{fuj05} and \cite{zeg07}, respectively. These
permit to estimate the EC lifetimes of a possible back decay to
$^{138}$Ba.

\begin{figure}[tbh]
\includegraphics[width=8.5cm]{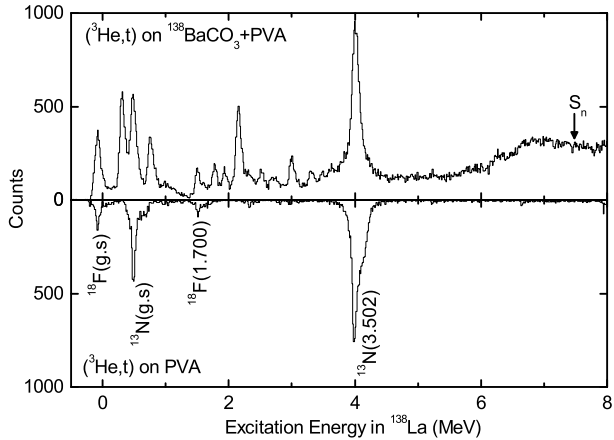} 
\caption{\label{fig1} 
Top: Spectrum of the $^{138}$Ba($^{3}$He,$t$)$^{138}$La reaction at
$E_0 = 420$ MeV and $\Theta = 0^\circ - 0.5^\circ$, taken from
Ref.~\cite{bye07a}.  The target consisted of $^{138}$BaCO$_3$
dissolved in PVA. Bottom: Spectrum of the ($^{3}$He,$t$) reaction on
PVA.}
\end{figure}

A spectrum of the reaction is shown in the upper part of
Fig.~\ref{fig1}, taken from Ref.~\cite{bye07a}. Since the $^{138}$Ba
target material was embedded in polyvinylalcohol (PVA), a background
line from the $^{18}$O($^3$He,$t$)$^{18}$F reaction is visible close
to the expected g.s.\ energy of $^{138}$La. The contribution of the
PVA was subtracted by a measurement on a pure PVA target under
identical kinematics (lower part of Fig.~\ref{fig1}).  No transition
was observed in $^{138}$La for energies between the g.s.\ and 72
keV. A conservative upper limit from the present experiment for the
population of a hypothetic low-energy state can be extracted varying
between $B({\rm GT}) = 0.04$ close to the g.s., where the background
line is prominent, and $B({\rm GT}) = 0.02$ around $E_x = 72$ keV.
It is used to estimate an upper limit of the corresponding half life
from the relation between $ft$ and $B({\rm GT})$ values (see, e.g.,
Ref.~\cite{hax95}).
%
%
For $^{138}$La EC decay one obtains upper limits ranging from 3.48 h
at $E_x = 72$ keV to 1.90 h at $E_x = 0$ keV.

On the other hand, excitation of the previously known $1^+$ state in
$^{138}$La at $E_x =293$ keV was prominently observed
(cf.~Fig.~\ref{fig1}) with $B({\rm GT}) = 0.42$. The exact $B({\rm
GT})$ values depend on the model used for conversion of experimental
cross sections to transition strengths, but differences between the
approaches of \cite{fuj05} and \cite{zeg07} are of the order of 10\%
in heavy nuclei only, which is of no relevance to the present
discussion.

The competing electromagnetic transition of a hypothetic low-energy
$1^+$ state to the $^{138}$La g.s.\ would be of $E4/M5$ character.
Contributions of $M5$ should be suppressed by several orders of
magnitude and are therefore neglected. The structure of the
hypothetical $1^+$ state is unknown; for an estimate we assume an
$E4$ transition strength of 1 Weisskopf unit. Because of the large
atomic number and small excitation energy the decay would be
dominated by internal conversion (IC). IC coefficients were
calculated with the code BRICC \cite{kib08}. The resulting half life
at $E_x = 72$ keV of a hypothetical state would be 8.9 d (and even
larger for lower $E_x$), still significantly longer than the limit
deduced for the EC decay.

Nevertheless, the experimental results \cite{bye07b} provide an
indirect argument against the existence of another low-lying $1^+$
state besides the lowest known one at $E_x = 293$ keV. The structure
of the lowest states in $^{138}$La can be understood in the simplest
approximation as proton-particle, neutron-hole states with respect
to the $N = 82$ closed-shell nucleus $^{138}$Ba. The single-particle
energies of shells near the Fermi level can be estimated from
single-nucleon transfer reactions populating $^{137}$Ba and
$^{139}$La, respectively. The lowest neutron-hole states observed in
$^{137}$Ba \cite{cha71} are $2d_{3/2}$ (g.s.), $3s_{1/2}$ ($E_x =
0.28$ MeV), and $1h_{11/2}$ ($E_x = 0.66$ MeV). For the
proton-particle states in $^{139}$La \cite{wil71} one finds
$1g_{7/2}$ (g.s.), $2d_{5/2}$ ($E_x = 0.17$ MeV), $3s_{1/2}$ ($E_x =
1.21$ MeV), and $1h_{11/2}$ ($E_x = 1.44$ MeV). A clear energy gap
is observed in both cases suggesting the lowest states in $^{138}$La
to be formed by the configurations $(\pi g_{7/2} \nu
d_{3/2}^{-1})_{2^+,3^+,4^+,5^+}$, $(\pi g_{7/2} \nu
s_{1/2}^{-1})_{3^+,4^+}$, $(\pi d_{5/2} \nu
d_{3/2}^{-1})_{1^+,2^+,3^+,4^+}$, and $(\pi d_{5/2} \nu
s_{1/2}^{-1})_{2^+,3^+}$. Thus, the low-energy spectrum of
$^{138}$La should consist of the following number of states with a
given spin $J^\pi =1^+$(1), $J^\pi =2^+$(3), $J^\pi =3^+$(4), $J^\pi
=4^+$(3), $J^\pi =5^+$(1). This is exactly what is found for the
lowest states in $^{138}$La up to an excitation energy of $E_x =
642$ keV \cite{son03}. The next-higher states all show negative
parity (where known) indicating that their structure involves the
$1h_{11/2}$ orbital.

The experimental observation of only one $1^+$ state at 293 keV with
a rather large GT strength is qualitatively consistent with the
above picture of a dominant $(\pi d_{5/2} \nu d_{3/2}^{-1})_{1^+}$
structure. Any further hypothetical $1^+$ state in $^{138}$La at low
excitation energies should mix with this one leading to a finite GT
strength which can be largely excluded from the sensitivity limits
of the ($^3$He,$t$) data. As an example, in a two-state model
assuming an interaction matrix element $V = 10$ keV and an energy
spacing $\Delta E = 250$ keV between the two states one would obtain
complete mixing with a corresponding share of the GT strength.

As a final remark, evidence against the existence of an intruder
$1^+$ state in $^{138}$La as the first excited state is also
provided by a study of the $^{138}$Ba($p,n\gamma$) reaction
\cite{isl75}. In this reaction low-spin states are preferentially
excited but no $\gamma$ transitions consistent with such a picture
were found. We conclude that the existence of a low-energy $1^+$
state in $^{138}$La which would affect the cosmochronometer
discussed in \cite{hay08} and also the analysis of charged-current
reactions as a major nucleosynthesis source of $^{138}$La
\cite{heg05,bye07a} is extremely unlikely.

\begin{acknowledgments}
This work was supported by the DFG under contracts SFB 634 and 446
JAP-113/267/0-2.
\end{acknowledgments}

\end{document}